\documentstyle[psfig,secnumtab]{fbssuppl}

\title{The Strange Magnetic Moment of the Proton
in the Chiral Quark Model}
\author{D. O. Riska\thanks{{\it E-mail address:} 
riska@pcu.helsinki.fi}}
\institute{Department of Physics, 00014 University of Helsinki,
Finland}

\begin{document}

\maketitle
\begin{abstract}
	The strange magnetic moment of the proton is small in the chiral 
quark model, because of a near cancellation between the quantum 
fluctuations that involve kaons and $s$-quarks and loops that involve 
radiative transitions between strange vector mesons and kaons.
 \end{abstract}

\centerline{hep-ph/9808381}

 \section{Introduction}
	The observation that the
strangeness magnetic moment of the proton may be positive
\cite{SAMP} ($G_M^s(Q^2=0.1$ GeV$^2)=0.23\pm 0.37$)
was unexpected as most
predictions had given negative values \cite{Mus94,Beis,Barz,Dong}.
The expectation of a negative value for
$G_M^s(0)$ may be explained as follows: 
Consider a $u$ quark with spin $s_z$ projection 
$+{1\over 2}$. If this fluctuates into a kaon and an $s$ quark,
the probability that $s_z$ of the $s$ quark be $-{1\over 2}$ is
twice that for the value $+{1\over 2}$. Hence, as the charge of the
$s$ quark is $-e/3$, the $s$ quark should contribute a positive
amount to the magnetic moment of the $u$ quark.
On the other hand following spin-flip the $\bar s$ quark in the 
kaon has
orbital angular momentum $l_z=+1$, and thus as its
charge is $+e/3$, it should also contribute a
positive amount to the net magnetic moment of the $u$ quark.
As there are twice as many $u$ quarks with $s_z=+{1\over 2}$
as with $s_z=-{1\over 2}$ in the proton it follows that
this fluctuation will give a positive contribution
to the magnetic moment of the proton.
As $G_M^s$ is defined as the matrix element 
of $\bar s\gamma^\mu s$
this matrix element is obtained by 
multiplication of the ``conventional''
magnetic moment by $-3$, and thus should be negative.

	A positive value for $G_M^s(0)$
has to arise from transition couplings between strange mesons.
It has recently been noted that a fluctuation into
a loop with a $K^*$ meson , which decays
into  kaon that is reabsorbed with the $s$ quark gives a
positive value \cite{Gei}. That work based on
the nonrelativistic oscillator quark model gave the
result $G_M^s(0)$ = 0.035. It is shown here that 
the chiral quark model leads to a small negative
value for $G_M^s(0)$. 

This calculation should be contrasted with those
in refs.\cite{Mus94b,Barz}, where the 
strangeness fluctuations were considered at the hadronic
level. The consideration of the strangeness fluctuations
of the constituent quarks 
is motivated by the fact that the magnetic
moments of the nucleons are explained by
the constituent quark model, but not by the hadronic
constituent model
\cite{Bethe}. The smaller
meson-quark coupling constants imply loop
corrections, which are sufficiently small so as
not perturb the overall quark model
description of the magnetic moments.
The quark model approach automatically
takes into account all baryonic intermediate states.

 \section{Strangeness Loops with Transition Couplings}
 \label{secstyle}
	Consider the (hidden) strangeness fluctuations $q\rightarrow 
K^*s\rightarrow K\gamma s\rightarrow q$ of a light $u$ or 
$d$ quark $q$. The key vertex in this loop diagram
is the $K^*\rightarrow K\gamma$ vertex, which is described
by the transverse current matrix element:
$$<K^a(k')|J_\mu|K_\sigma^{*b}(k)>=-i{g_{K^*K\gamma}\over m_K^*}
\epsilon_{\mu\lambda\nu\sigma} k_\lambda k_\nu^{'} 
\delta^{ab}.\eqno (1) $$
The coupling constant $g_{K^*K\gamma}$ is determined
by the empirical decay widths for radiative decay of the 
$K^*$ as
$g_{K^{*+}K^+\gamma}=0.75$ and $g_{K^{*-}K^-\gamma}=1.14$.
 
Given the current matrix element (1) the contributions to the
anomalous magnetic moment of the $u$ and $d$ quarks from the
$K^*K\gamma$ loop diagrams take the following
form (when expressed in terms of nuclear magnetons and after
assigning the kaon line a ``strangeness charge'' of $-1$):
$$G_M^s(0)_{u,d}=-{g_{Kqs}g_{K^*qs}g_{K^*K\gamma}\over
2\pi^2} {m_p\over m_K}
\int_0^1 dx(1-x)\int_0^1 dy \lbrace m_u(m_s-m_u x)$$
$$\lbrace{1\over G_1}-{1\over G_2}-{1\over G_3}+{1\over G_4}\rbrace
-log({G_2 G_3\over G_1 G_4})\rbrace .\eqno(2)$$
Here $m_p$ is the proton mass and $m_u$ and $m_s$ the 
constituent masses
of the $u,d$ and $s$ quarks respectively for, 
which use $m_u=$ 340 MeV, $m_d=$ 460 MeV \cite{GloRis}. 
The quantities $G_j$ above are
defined as
$$G_1=G(m_K,m_{K^*}),\, G_2=G(\Lambda,m_{K^*}),\,
G_3=G(m_K,\Lambda^*),\, G_4=G(\Lambda,\Lambda^*), \eqno(3)$$
where the function $G$ is defined as
$$G(m,m')=m_s^2(1-x)-m_u^2x(1-x)+m^2x(1-y)+m^{'2}xy.\eqno(4)$$
In the case of the $u$ quark the coupling $g_{K^*K\gamma}$
constant is that for the $K^+,K^{*+}$
mesons and in the case of the $d$ quark it is that for the
$K^0,K^{*0}$ mesons.

The $\pi -$quark coupling constant is
determined by the $\pi NN$ coupling
$g_{\pi NN}$ as \cite{GloRis}:
$g_{Kqs}=g_{\pi qq}=3/5(m_q/m_N)
g_{\pi NN}=2.9$.

The corresponding vector meson coupling constant $g_{K^*qs}$
is related to the
nonstrange vector meson ($\rho$) quark coupling strength
as
$g_{K^*qs}\simeq g_{\rho qq}=(3/5)(m_q/m_N)
g_{\rho NN}$.
With the usual value for the $\rho NN$ vector coupling
constant $g_{\rho NN}^2/4\pi = 0.52$ and the including the
factor $(1+\kappa)$, where $\kappa=6.6$ (\cite{Hoe}) is the
$\rho NN$ tensor coupling, the value $g_{K^*qs}=g_{\rho qq}
=4.2$ obtains.

As the $K^*Ks$ loop diagrams are logarithmically divergent a
cut-off is included in their evaluation
by insertion of monopole form factors of the form
$(\Lambda^2-m_K^2)/(\Lambda^2+k^{'2})$ and
$(\Lambda^{*2}-m_K^{*2})/(\Lambda^{2*}+k^{2})$ at the $K$ and $K^*$
vertices respectively. The cut-off masses $\Lambda$ and
$\Lambda^*$ are free parameters. We shall here take $\Lambda$ =
$\Lambda^*$ in the numerical calculations.

The contribution from these loops to the strange magnetic
moment of the proton is obtained as
$$G_M^s(0)_p={4\over 3}G_M^s(0)_u-{1\over 3}G_M^s(0)_d,\eqno (5)$$
and thus will have a value close to the corresponding
constituent quark values.
For $\Lambda$ values in the range 1--1.5 GeV the 
contributions are positive, but small (Table 1).

	These values are smaller than the
corresponding value obtained in ref.\cite{Gei} with
the harmonic oscillator quark model. The difference is
due to the nonrelativistic nature of the latter which
implies an overestimate of the radiative widths
of the vector mesons. With the usual static fermion 
currents of the constituent quark model the coupling
constant $g_{K^*K\gamma}$ should be 2, which is 
twice too large. 

\centerline{\bf Table 1}

 Contributions to the strange magnetic moment of the proton
(in nuclear magnetons) 
from the loops with $K^*\rightarrow K\gamma$ transition
vertices and with diagonal $KK\gamma$ vertices
as a function of the cut-off parameter
$\Lambda$ (in MeV).

\begin{center}
\begin{tabular}{|r|l|l|l|} 
\hline
$\Lambda$ & $K^*K\gamma$  & $KK\gamma$ & Sum\\
\hline
800  & 0.0   &-0.016 &-0.016\\
1000 & 0.006 &-0.031 &-0.025\\
1200 & 0.026 &-0.046 &-0.020\\
\hline
\end{tabular}
\end{center}

\section{Strange Fluctuations with Diagonal Couplings}

The expressions for 
the contributions from the diagonal $Ks$ loop contribution 
to the strange magnetic
moments of the $u$ and $d$ quarks 
are similar to the corresponding expressions for the 
contribution of pionic
fluctuation to the neutron magnetic moment \cite{Bethe}. 
These kaonic loops then give the same 
contribution to the anomalous strange magnetic moment of the 
$u$ and $d$ quarks. This may be written in the form
$$G_M^s(0)=-{g_{Kqs}^2\over 4\pi^2}m_p\int_0^1 dx(1-x) (m_s-m_u x)
\{(1-x)[{1\over H_1}-{1\over H_4}-x{(\Lambda^2-m_K^2)
\over H_4^2}]$$
$$+\int_0^1dy x[{1\over H_1}-{1\over H_2}-{1\over H_3}
+{1\over H_4}]\},\eqno(6)$$
Here the quantities $H_j$ are defined as
$$H_1=H(m_K,m_K),\, H_2=H(m_K,\Lambda),\, H_3=H(\Lambda,m_K),
\, H_4=H(\Lambda,\Lambda),$$
$$H(m,m')=(m_s^2-m_u^2 x)(1-x)+m^2xy+m^{'2}x(1-y).\eqno(7)$$

By (5) the kaonic loop contribution (6) to the
constituent quark equals that of the proton. For 
$\Lambda$ below 1.2 GeV this negative contribution is larger
in magnitude than the positive contribution of the 
$K^*Ks$ loop contributions considered above.

 \section{Discussion}

The result obtained here is that the strange magnetic moment
of the proton is small because of a strong cancellation between
strangeness fluctuations, which involve radiative transition
couplings and such which do not. The chiral quark
model calculation leads to a stronger cancellation than the
non-relativistic quark model calculation in ref.\cite{Gei}.

\vspace{1cm}

\centerline {\bf Acknowledgments}

This
work was supported in part by the Academy of Finland under 
contract 34081 and is based on a collaboration with
L. Ya. Glozman.

 \SaveFinalPage
 \end{document}